\documentclass[12pt]{article}
\usepackage{amssymb}

\newcommand{\bo}{{\bar o}}

\def\bo{{\raise.15ex\hbox{\large$\Box$}}}               

\def\face{{\raise.2ex\hbox{$\displaystyle \bigodot$}\mskip-2.2mu \llap {$\ddot
        \smile$}}}                                      

\def\leftrightarrowfill{$\mathsurround=0pt \mathord\leftarrow \mkern-6mu
        \cleaders\hbox{$\mkern-2mu \mathord- \mkern-2mu$}\hfill
        \mkern-6mu \mathord\rightarrow$}       
\def\dvec#1{\vbox{\ialign{##\crcr
        \leftrightarrowfill\crcr\noalign{\kern-1pt\nointerlineskip}
        $\hfil\displaystyle{#1}\hfil$\crcr}}}           

\def\beq{\begin{equation}}
\def\eeq{\end{equation}}

\def\beqx{\begin{displaymath}}
\def\eeqx{\end{displaymath}}

\def\beql{\begin{eqnarray}}
\def\eeql{\end{eqnarray}}

\newcommand{\bea}{\begin{eqnarray}}
\newcommand{\eea}{\end{eqnarray}}

\def\[{\left [}
\def\]{\right ]}
\def\({\left (}
\def\){\right )}

\def\+{\oplus}

\begin{document}

\vspace*{0.15in}
\hbox{\hskip 12cm NIKHEF/2006-004  \hfil}
\hbox{\hskip 12cm physics/0604134 \hfil}
\begin{center}
{\Large \bf The Landscape ``avant la lettre"\\}

\vspace*{0.5in}
{ A.N. Schellekens${\,}$\footnote{t58@nikhef.nl}}\\[.3in]
{
         {\em NIKHEF Theory Group \\
	Kruislaan 409 \\
	1098 SJ Amsterdam \\
	The Netherlands \\}
	~~~~~\\
	and\\
	~~~~~~\\
	{\em IMAPP, Radboud University, Nijmegen\\ }
	}

\end{center}

\begin{center}
\vspace*{0.20in}
{\bf Abstract}

\end{center}

This is a translation of an inaugural speech given originally
in Dutch in 1998. The topic of that speech, intended for a general
audience,  was what is now called
``The Anthropic Landscape of String Theory".

\newpage

\section{Introduction}   

In 2003 L. Susskind published his paper entitled
 ``The Anthropic Landscape of String Theory" \cite{ALS}, which
I read with great pleasure. The reason was that,
many years before, I had come to the conclusion 
that everything we knew about String Theory was pointing towards an
``anthropic landscape" of vacua. I had advocated this idea consistently during
many years, on the basis of far less evidence than we have today. It seemed
obvious to me, but the response I got was frustrating.
Therefore I was delighted 
that finally someone was
stating this point of view loud and clear.
As Susskind correctly points out, the idea that the standard model
might not be the unique outcome of String Theory
seemed unacceptable
to most people. In seminars he gave on the subject that statement
was phrased more strongly, and became something like
``Until
 recently all string theorists believed that the laws of our
 universe will follow uniquely from some selection
 principle in string theory".  
I am quoting from memory, but I am certain
 about the ``all". Although I basically agree with that statement,
 the word ``all" is an exaggeration.
 
Of course I understand that the expression
 ``all string theorist" was used for theatrical purposes, and
that the author must have known that at least a few string theorists have thought
 and talked about a possible
 non-uniqueness of the outcome, and about a possible
 r\^ole of the anthropic principle in relation to the string vacuum
 problem. For example, there is a footnote in a review
article by F. Quevedo \cite{Quevedo} from 1996,
which discusses the anthropic outcome at least as an option
that should be considered seriously.    
However it does seem to be true that remarkably
 few people have ever put their thoughts on this subject in
 print\footnote{I do not even know any
other early examples than the footnote in \cite{Quevedo}.  However, such statements are 
hard to search for. I would be very much interested in learning about similar
ones.}. That is a pity, because there are many subtle distinctions between various
possible points of view, and it would have been interesting to know what other people
were really thinking at a particular moment.

 The purpose of this paper is to publish a text which may well be
one of the earliest ones advocating the ``Anthropic Landscape", as it is now called, unequivocally.
 It is the translation of a 
 speech I gave in 1998 \cite{DutchText}, in Dutch, on the occasion of my
 inauguration at the University of Nijmegen.
 The Anthropic Landscape is not just described in this
 speech, it is the main topic. I tried to argue that an Anthropic Landscape
 of ground states of a consistent fundamental theory
 is not just the most likely
 but also the most desirable outcome of the age-old quest
 towards understanding our universe. 
 In this respect, my arguments went quite a bit further
 than any other texts from that period that I am aware of.

I am not posting this text to claim any part of the fantastic discoveries made
during recent years (for example \cite{Bousso:2000xa}, \cite{Kachru:2003aw}; see 
 \cite{Polchinski} for a review and a
complete list of citations). Indeed, the text says nothing about flux compactifications
or the cosmological constant. I am posting it for a number of other
reasons. I want to provide concrete evidence that rather precisely
formulated ideas of this kind existed, and in written form,
well before 2003.
I also want to
demonstrate that it was possible to arrive at this point of view even before the
recent discoveries were made. 
Furthermore I have been fascinated by the reaction to these ideas
for many years now, and I think even today we are only in an intermediate
stage of a very slow shift of opinions regarding the objectives of our field. Although
landscape ideas and even the anthropic principle are now at least discussed, it seems
to me that  
 the importance of the landscape is still severely underrated.
I have tried to express my enthusiasm about the recent
progress during seminars, but apparently with little success. I think my
feeling of excitement is a lot easier to understand  if I  back it up with this text,  which precedes all recent
developments, and hence was not influenced by them. If indeed the landscape gets the
important place in science history that I expect, historians of science may be very puzzled
why it took string theorists so long to arrive at this conclusion. I hope this text will help
in answering that question.
 
 My own thoughts in this direction started around 1987. The
 year before I had published a paper with Wolfgang Lerche  and Dieter L\"ust \cite{Lerche}.
 Like other authors at the time, we found large numbers of four-dimensional chiral
 string theories, but much more than others
 we made a point of strongly emphasizing
 the non-uniqueness of the result. In the following year we began to understand that this
 was perhaps a bit premature. There were moduli to be fixed, and supersymmetry
 had to  broken, and of course that could drastically change the conclusion.
 But it seemed to me that it was wishful thinking to assume that all these
 problems would be solvable for just one ground state, the one corresponding
 to the standard model. 
 In addition, the standard model did not look like a unique solution to anything,
and seemed fine-tuned to our own existence. I came to the conclusion that String Theory
itself suggested the obvious way out of the dilemma of having at the same time a unique
theory and on the other hand something as special as the standard model: a unique theory
with a huge number of vacua. A r\^ole for the anthropic principle was inevitable in this
scenario. A unique outcome of any argument not involving our own existence would create
a disastrous anthropic fine tuning problem. Inexplicably, a mathematical computation would
yield a point precisely in the presumably extremely small
 region of parameter space allowing intelligent life.

It seemed obvious. As obvious as saying that not the earth but the sun
was the centre of the solar system. I expected that everyone would agree
 with me very quickly. But I was surprised by the reactions I got
when I brought this up. I learned that the word ``anthropic" provoked strong
and often irrational reactions. And that there was indeed, as mentioned above,
a strong hope or belief in an ultimate unique answer. In fact I say so
in the 1998 text.

I have considered many times writing a paper about my ideas, but I could not bring myself
to write something that seemed so obvious. It would have been a paper with many
words an no formulas, and it was not at all clear where to publish it. There were
no blogs, no homepages, no ``arxives", and no obvious journal to send it to. 
 And furthermore the hard evidence was not
available yet. 
One valid objection was that we did not know enough about string 
theory yet to make any claim about the existence of a large number of non-supersymmetric, stable vacua. Today
that is still an often-heard objection.
My feeling about that was exactly what was written in \cite{ALS},
 ``if we find one such vacuum we are going to find a huge number of them". To me that
was clearly the message String Theory was sending us already in 1986, but most people
preferred to ignore it, although of course everyone agreed that the number of {\it supersymmetric} vacua
was huge.

It is difficult and dangerous to claim with the benefit of hindsight that one arrived at
some conclusion at a particular time. After 1986 it took some time to understand that 
all four-dimensional string vacua that were proliferating quickly had moduli, and that this
was an important problem. It took some time to appreciate that they were all related to each other,
and could be thought of as ground states of one theory, the Heterotic string. And of course there
were five string theories, not just one. The most common attitude was to ignore the others
and assume that one day we might know what was wrong with them. 
This was also my point of view until 1994, but
uniqueness of the underlying theory was anyhow not the most crucial part of the argument. Apart from that
issue, I was  already defending a point of view quite similar to the one expressed in the 1998 speech during
my time at CERN, which in any case means before 1992. I had discussions about that with many
people and encountered a lot of resistance, and I do not recall anyone wholeheartedly agreeing with me. 
It is therefore somewhat strange that
after 2003 some people started telling me ``this is what I have always been saying".

I would be one of those people now if I had not been offered a unique opportunity to put
my point of view on record. I was put under some gentle pressure to give a speech on
the occasion of my inauguration in Nijmegen. Usually such speeches describe the status of the
field, with some emphasis on glorious achievements of theory and/or experiments, and then give
some outlook on the future, or some comments on a
theme of particular
interest to the speaker. I was not particularly keen on giving such a speech until I realized that
this, at last, was a unique opportunity to publicly advocate the point of view I had been talking about
for so many years.
Furthermore the simplicity of the main point was only a bonus here. These talks are typically attended
by members of the family, friends, and colleagues from the same and other universities. This requires
a difficult balancing act in order to explain something to an audience of outsiders of the field, and
also trying to tell something worthwhile to your colleagues. The main point was simple to explain and
at the same time controversial enough to be of interest.  

The text of these speeches is printed as a small booklet
and distributed among staff members and libraries of the various institutes in the Netherlands. The Dutch text is
therefore authentic and unchangeable. It is included in this paper so that Dutch speaking
persons can verify that I did not use hindsight to attempt to make changes during the translation, which was made in 2006.

There were in fact not many temptations to make changes. Reading the text again after
eight years, 
I was worried to encounter remarks about which
I had changed my mind in the meantime, but I was pleased to see 
that most things were written carefully, exactly
as I often state them today, and exactly as I remember stating them at CERN in the late eighties. I was
amused to discover that I even used the word ``landscape" in the text, or rather the Dutch
word ``landschap" from which the English word was derived in the 16th century. That word was used once,
in a sense not very different from the current one, but the expression I really used for what would
now be called the ``landscape of string vacua" was the Dutch word ``gebergte", which translates as ``mountain range". 
If I had translated this text to English in 1998, I would certainly have tried to find a more attractive
English term, and ``landscape" would have suited my purpose very well. But of course that's just a name.

The Dutch title was ``naar een waardig slot", which means something like ``towards a
worthy end". I was trying to think of possible ways in which the story of high energy
physics might end in a way worthy of its wonderful history,
 and concluded that a unique theory with a huge number of vacua and anthropic
{\it raison d'\^etre} of our universe was the most attractive outcome, and that this was precisely
what String Theory was suggesting.

When reading the text one should keep in mind that it was written for a general audience. This required the
usual shortcuts and simplifications. It also required me to drop some important issues. It made no sense
to try and explain the problems of moduli stabilization or supersymmetry breaking in this context. The only
place where I express
my worries about that is in the sentence ``At this moment it is not clear what 
will be left of this enormous number of ground states once we understand
String Theory properly". The cosmological constant is another important topic that is not mentioned, partly
for the same reason, but mostly because at that time I was seriously confused about it. In 1987, when
I started thinking along these lines, I assumed it was zero, and that we would find some mechanism to explain that.
By 1998 evidence was coming in that it might not be zero, and I had become aware of Weinberg's ``anthropic"
argument \cite{Weinberg}. It looked like the cosmological constant was the anthropic variable {\it par excellence}, but this
fact did not seem to fit in with the rest. I was convinced that a huge number of vacua 
was needed to understand
the anthropic fine-tunings of the standard model, but   
to explain the 120 orders of magnitude fine tuning of the cosmological constant seemed too absurd to even
consider. Nothing I knew suggested that there could be so many vacua.

Indeed, the most important point on which I was forced to change my mind since
 1998 has been the size of the landscape. There is no
clear statement about how large I expected the landscape to be, but the text does give a hint about that.
At that time I was convinced that the current experimental data would be sufficient to pinpoint one
particular vacuum as the one corresponding to our universe and our standard model.
Then, if we would just find that one, everything else could be predicted. 
According to
recent rough estimates by M. Douglas, that would be the case if there are less than about $10^{80}$ vacua
(more or less uniformly distributed over a bounded region of parameter space), and if
one does not take tuning of the cosmological constant into account.  The amount of anthropic tuning in
the standard model parameters is essentially impossible to estimate at present with any accuracy,
because it requires computing all of cosmology, star evolution, chemistry, nuclear physics and biology etc. for
a different choice of gauge group, representations and couplings. It is a number between $10^{-80}$ and 1, and presumably
far from either limit. If the result were smaller than $10^{-80}$ we could use the anthropic principle to
improve on the current experimental bounds on the standard model parameters, and this seems ridiculous.
If the result were 1 this would imply that any choice of gauge theory, representations and
parameters leads to something one could call ``life". If that notion
seems too ill-defined for generic universes one could drop the ``biology", and
talk about complex chemistry instead. 
It seems unlikely that
a randomly chosen theory would lead to sufficient chemical complexity,  but, as I say in the
text, that is an assumption. It is clear that the entire
argument breaks down if the probability approaches 1.
If indeed it is much smaller than 1,
this kind of argument would require the
number of vacua to be at least the inverse of that probability, and considerably larger to avoid any
illusion of fine-tuning. In 1998 I was more worried that the number of string vacua that I knew was too
small, than that it would be far larger than this lower bound.

Note that in the 1998 text
I did not advocate the idea that any particular standard model parameter can be determined from
anthropic arguments alone. An often stated objection to the anthropic principle is that it is
like giving up physics and just declaring instead that ``things are the way they are, since otherwise
we would not exist". However, as I already explained in 1998, the proper use of this principle presupposes
a precise understanding of the landscape. It may well be that the anthropic principle is of no
practical use for the determination of any standard model or cosmological parameter with the
possible exception of the cosmological constant, simply because the necessary computations are
too hard.

There will certainly exist functions on the landscape that affect the vacuum selection, such as
the local density of certain kinds of vacua \cite{Denef:2004ze}
or cosmological likelihood. It is important to compute
such quantities, if possible, but I am convinced that they
will not uniquely select our universe, for exactly the same reasons I stated in 1998. In fact, neither
will the anthropic principle, which is just another function on the landscape (although hard to define precisely), giving the
likelihood that ``intelligent" life emerges in a given point on the landscape (an analogy with the
``Drake equation" suggests itself here rather naturally). It would be absurd if our universe were the only
point in the landscape that is sufficiently anthropic, and nothing we know suggests that.
Perhaps some sort of convolution of all these
functions will provide an answer close to the Standard Model.

Most people associate the Landscape with an anthropic solution to the cosmological constant problem.
For me that was not the crucial issue. Even if 
we came to the conclusion that the cosmological constant was not anthropically tuned after all, I would still
expect an anthropic landscape for the structure and parameters of the standard model. String Theory was
the first and so far only theory that made the question about uniqueness of the 
standard model unavoidable. Most, if not all, other attempts
to ``derive"  the structure of the Standard Model involve a new layer of gauge theories, for example
composite models or GUTs. Then on inevitably runs once again into the same problem one tries to solve, namely
an essential non-uniqueness.  The most promising candidate, $SO(10)$ Grand Unification, requires an
unexplained triplication of families and has a large number of parameters, even if one takes this choice
of gauge group for granted.  The only hope for uniqueness is a theory that itself has a chance of being
unique, namely a theory of gravity.
Such a theory, String Theory, was explored during the past decades, and it gives a very clear answer:
there is no unique ground state, but a landscape of vacua. It is in my opinion 
the only answer that makes sense, and the fact that this answer came out of String Theory is a sign that
we are on the right track. 
I see this as a fundamental result that may even
survive if String Theory turns out to be incorrect, or if String Theory is just the tip of an iceberg.

I expect that the String Theory Landscape will acquire an important place in science history. 
Of course its ultimate fate depends on the correctness of String Theory, and the
unexpectedly huge size of the landscape is making it a lot harder to
convince ourselves of that. But String Theory won't
be correct without the landscape being correct. And if that is true, it would be one of the most fundamental
discoveries one can make. It implies that we would know the answer to Einstein's question if the creator of
our Universe had any choice: indeed, we would know all the choices.   
This insight is probably the most important one we have obtained from string theory so far. It should
be remembered that in 1984 this would have been completely unthinkable. Unlike the other main result we 
hope to get out of String Theory, consistent Quantum Gravity, the landscape emerged against everyone's initial
expectations and wishes. It is a revolution that is unfolding so slowly that few people even recognize
it as such. But nevertheless, discovering that our standard model is just one entity in a huge landscape,
and hence cannot be completely derived from first principles, is a paradigm shift for our field.

I am tempted to continue with further remarks on the present state of affairs, and comment on some
of the papers and other texts that have appeared in the last few years. The fact that these issues are
now being discussed is by itself an exciting development.
But many of the remarks I could make would
just be a rephrasing of what I wrote in 1998, and I prefer to let that text speak for itself.

\section{The English Translation of \cite{DutchText}}

``High Energy Physics" has a special place within physics.
This is already apparent from its name. Most fields within
physics are called after what they study. Solid state physics
studies properties of solids, atomic physics studies atoms,
and nuclear physics nuclei. But a high energy physicist does
not study ``high energy". The name of the field is derived from
the experimental techniques that are used. The essence of
this field is the study of matter at ever shorter distance scales.
To achieve this, beams of particles of very high energy are collided
against each other. The higher the energy, the deeper one penetrates into
the matter. Our field effectively started with the invention of the
microscope, but it really started moving ahead in this century. We
learned that matter consists of molecules, which are built out of
atoms. The atoms consist of a nucleus with a cloud of electrons around it,
and the nucleus consists of protons and neutrons. Every time you look
deeper, new phenomena seem to appear. Perhaps this is the reason our field
is not named after its field of study; after all, this changes all the
time. Sometimes one calls it ``elementary particle physics", but when that
name was introduced  one believed that protons and neutrons were
elementary. Meanwhile we know that this is not true. The proton
and the neutron are not elementary, but consist of other particles,
the quarks. If it turned out that the quarks themselves consisted
of other particles, our attention would automatically shift to those
particles. There is, however, no evidence for this at present.

One could say that the essential difference between High Energy Physics
and all other branches of physics is that the former investigates
vertically, whereas all other investigate horizontally. That definition
essentially implies an end of the discipline. It seems inevitable
that one day we will run into a limit which makes it impossible
to penetrate deeper into matter. This would imply that nature at its
deepest level will forever be beyond our observation. If what we
cannot see is as complicated as what we have seen already, there
will be no chance to guess the answer. This seems the most likely
end of our field.

It is less clear when that end would be reached. During the
last 50 years the technological limits have been shifted continuously,
and more and more impressive machines were built. One of these
successes is for example LEP in Geneva, a circular particle
accelerator with a circumference of 27 kilometers, where electron and
positron beams are collided with large energy and astonishing
precision. Each time such a machine is commissioned, the next one
is already on the drawing board. Technologically the end may appear
to be in sight, but that is already true for twenty years, and there are
still new accelerators being planned or being built.

An unforeseeable technological revolution could postpone the end of
this road substantially, but eventually we will encounter fundamental
or practical limitations. This becomes painfully clear if we consider
the smallest distances that might be of interest theoretically. In theories
of Quantum Gravity interesting phenomena are happening at distances
the size of the Planck length, which is about $10^{-33}$ cm. This is so
terribly small that observing it with our most powerful particle
accelerators is something like trying to see protons on the moon with
a telescope. With an accelerator like LEP we can penetrate to about
one hundredth of the size of a proton. A machine that can see structures
the size of the Planck length has to accelerate particles to a gigantic
energy, the Planck energy. To construct such an accelerator with
current technology would require a machine with at least the size of
our entire galaxy. It is an interesting question if this can be done
even theoretically, but it can certainly not be done in practice. 

Although the majority of experiments is done with particle
accelerators, there are other, less direct methods to learn
something about the deeper structure of nature. But no matter
which method is used, one can always imagine phenomena which are
beyond our observation. We can imagine particles with so little
interaction, or events that are so improbable that we will never see
them. It is difficult to see why nature should always be so friendly to
us to let us see all its secrets.

I regard this as an unpleasant idea. I see the development of fundamental
physics in this century as an exciting book, which is difficult to put
aside without having read the last chapter. It is a beautiful story that
deserves a worthy end. It would be very disappointing if the last chapter
turned out to be forever unreadable to us. 

Then what is really our goal? Most of my colleagues will only formulate
an answer to this question with great hesitation, because it sounds on the
one hand rather arrogant, and on the other hand hopelessly naive. You
could describe it as the ``Theory of Everything". This term was used for
the first time a few years ago in a moment of euphoria about a new breakthrough,
and was generally met with derision. What do you mean, ``Theory of everything"?
Does that theory have something to say about superconductivity? Can you derive
the human genetic code from it? Can you use it to predict the weather? The list
of questions humans struggle with is endless. A theory that answers all questions
is unthinkable.

If we nevertheless want to formulate a reachable and acceptable final goal, we
will have to limit the questions to which we may expect an answer. This implies,
among other things, that the questions should not be too complicated. For example
the DNA molecule is so complicated, that it will never be derived directly from
a fundamental theory. Nevertheless we know already for decades what that fundamental
theory is: it is Quantum Electrodynamics. When we ask sufficiently simple
questions, this theory gives extremely precise answers. In high energy physics most
questions are ultimately reduced to simple questions, such as: what exactly
happens when two particles collide with a given energy. When we have convinced
ourselves that all questions of that type can be answered one of the requirements
of a Theory of Everything has been met, as far as I am concerned.

In fact, we almost have such a theory at our disposal. It has been given
the not very impressive name ``Standard Model". This theory describes
three of the four fundamental forces we know.  In the first place this
includes the already mentioned Quantum Electrodynamics.  This interaction
manifests itself in many ways in the world around us: light, electricity,
magnetism, chemistry and most properties of materials are  all controlled
by this theory. Then there is  Quantum Chromodynamics, which among other things
is responsible for the fact that protons and neutrons are held together in
nuclei. Finally there is the weak force, which manifests itself
for example in radioactivity. For all these phenomena we have a very an
extremely successful  fundamental theory, that for already twenty-five
years passes all experimental tests convincingly. When people
look back at this century  in the future, the Standard Model will undoubtedly be
mentioned as one of the great breakthroughs in physics, together
with Quantum Mechanics and Relativity.

Nevertheless no-one will be inclined to replace the unpretentious name
``Standard Model" by ``Theory of Everything". There are a few reasons
for this. First of all one of the four known interactions, gravity, is
missing. At first sight this does not look like a  really big problem.
For gravity we also  have an excellent theory, Einstein's
General Relativity, one of the crown jewels of this century. Just like
the Standard Model, this theory makes extremely  precise predictions that agree
with a large number of experiments. Unfortunately  this theory does not
coexist very peacefully with Quantum Mechanics. This implies
concretely  that we are unable to make predictions when particles
collide with each other with Planckian energies, or for extreme
phenomena in the neighborhood of black holes. It is extremely unlikely
that we will ever be able to actually observe such processes, but we
expect a theory of everything to give an answer in these circumstances as well.

The Standard Model, on the other hand, has no problems with Quantum
Mechanics. It is formulated in terms of Quantum Field Theory, which is
founded solidly on its two pillars, Quantum Mechanics and Special Relativity.
Nevertheless  we are not  at all certain that the Standard Model
is correct. The Standard Model contains 17 particles, 16 of which have
been found so far. One is sill missing, the so-called Higgs particle.
This particle is rather different from the other 16, and is generally
seen as the weakest link in the Standard Model. The existence of this
particle was postulated twenty-five years ago, because it is
essential for the consistency of the Standard Model.  But on
the other hand it became clear fairly quickly that it introduces a new
problem. I will not explain in    detail what the problem is.
Essentially, the problem is that the weak force, despite its name, is
far stronger than one would reasonably expect.

There is more about that Standard Model that we don't like.  For
example it has a lot of parameters, namely at least 19  and probably more.
A parameter is a number whose value is not fixed by the theory. 
For example the Standard Model does not tell us what the quarks weigh.
The masses of the six quarks can only be determined by measuring them.
Once all parameters have been measured, the theory is fixed and every
other quantity can be computed. Although 19 parameters  is not really
very much, we might have preferred fewer, for example none at all.

What worries us even more is the fact that some of these parameters
have rather special values.  For example the ratio of the mass of
the electron with that of the heaviest  quark, the top quark
is very small, less than $10^{-5}$. Since the Standard Model does not
say anything about the masses, every number could have come out. When you observe
then that such a number  is  very small in nature, you get the feeling that you
should be able to understand that. The aforementioned problem
with the Higgs boson is of a similar nature, but more severe. It  is  like
being in a room with a hundred people, and discovering that they are
all born on the same day. The likelihood that this happens by
accident is so small that you would immediately assume that there has to be an
explanation. In the physics of our universe we see a variety of such
unlikely coincidences.

Something else many people don't like about the Standard Model is that
it is by no means the only possibility. When Einstein wrote down the
Theory of General Relativity he basically did not have any choice.
But the people who conceived the Standard Model had lots of variations
at their disposal. Thirty years of intense experimental research have 
thrown most of these  variations into the dustbin, but many people
would really prefer to do that on purely theoretical  grounds. They
would really prefer a unique theory.

Nevertheless it seems less certain
that we should be able to answer this type of question. Essentially we only have questions
of this type left in the Standard  Model. These are all ``why" questions.
Why exactly this Standard Model, why do the parameters have the observed
values and why are certain values so small. It is inevitable that some questions
of this type remain in a theory. The fact that such questions are not
answered is not a reason to withhold the predicate ``Theory of Everything".
We should definitely do that if it was inconsistent or in disagreement 
with experiment.

What I have been trying to say in the foregoing is that the concept
``Theory of Everything" is not necessarily a ``castle  in the sky"  provided
one does not have the wrong expectations. The Standard  Model is not
only an excellent model for three of the four forces of nature and all
matter, but also for the very concept ``Theory of Everything". Apart from  a 
few issues, it is really precisely what we are looking for.

The Standard Model has another problem that is rarely mentioned as such, but
that for me is essential. It is in complete agreement with everything we
can see, but it has nothing to say about what we cannot see. If new
experiments  discover new particles or interactions we will have to
extend the Standard Model. Quantum Field Theory allows lots of extensions.
We may postulate that nature is described by the Standard Model plus
nothing else, but nothing guarantees that this is true. We will always
depend on future experiments. This seems such an obvious point that 
no-one calls it a problem.

So what are our chances of ever finding that acceptable theory? I have
already said that I am not optimistic about that. The Standard Model
is not terribly complicated. If you use the correct
language you can write it down in two or three lines. Nevertheless
many experiments were needed to reach that point. Experiment and theory
need  each other. Without experimental help we have little chance.

But there may be a small chance.  At the moment a theory is
being developed that gives us some hope that perhaps we may accomplish this
task. I am referring to something we call ``String Theory", usually
translated to Dutch as ``Snaar  Theory", but I don't like the sound of that.
The theory gets its name from the fact that all particles  are realized as
vibrations of a kind of fundamental string. The lowest harmonics 
correspond to the particles of the Standard Model, plus perhaps a  
few new particles. The higher harmonics  correspond to an infinite
series of particles that we can never observe, unless  we can build a Planck Energy
accelerator. [....]\footnote{This phrase was not translated because it discusses
why I prefer the English name ``String Theory" over the Dutch name ``Snaar Theory", and
this explanation make no sense in English.} 

Meanwhile it has become clear that the word ``String" is much too limited and
the word ``Theory" a bit premature. We have discovered that in addition to strings
also membranes play a role, and we have realized that what we called ``String Theory"
until now is just a small piece of a continent that to a large extent still has to be
mapped out. For lack of a better name I will call that entire continent ``String Theory"
henceforth.

String Theory was discovered thirty years ago more or less accidentally in an
attempt to understand the strong interactions. It became clear rather
rapidly that this theory had magical properties, but soon a better
alternative was found for the strong force, namely Quantum Chromodynamics. During
ten years the theory, together with a handful of loyal followers, was essentially ignored, but
in 1984 it made a triumphant return.  In the mean time it had become clear that
the theory had promising features for the description of an entirely 
different force, namely gravity. A few enlightened persons had already understood
this in 1975, but they got little attention.

The 1984 revolution can be explained in particular by the fact that String Theory
offered us a gift nobody had counted on. Apart from a possibly consistent
theory of Quantum Gravity, String Theory turned out to contain particles and
forces that resembled those of the Standard Model. This is a generic feature
of good theories: bad theories always lack something essential, whereas good
theories give you lots of things you could not reasonably count on.

If everything works out the way we hope this should, in the end, give us a theory
that is completely consistent and contains all known particles and  forces. I
remarked before that we really would like a fundamental theory to give us also  a complete
description of everything we can not (yet) observe.
String Theory has a property that seems to realize that.
I am referring here to something
that in special cases is described by the technical term ``Modular Invariance", and
whose precise general form is not completely clear yet. In any case everything seems
to indicate that one cannot freely add particles to a given String Theory, or
remove them. Once we have found a String Theory that contains exactly the Standard
Model, we cannot add anything anymore. One cannot modify a string theory. The theory
may predict new particles, but if an experiment finds a particle that was not
expected, the entire structure collapses. Therefore String Theory can only yield
a theory of everything, whether we like that term or not: it is all or nothing.

The work ``nothing" is a bit exaggerated. At the very least we have a breathtaking
piece of mathematics in our hands. String theory attracts theorists because
it has marvelous features that are hard to explain to outsiders. It is a huge
landscape through which theorists move around with a kind of Alice-in-Wonderland feeling.
Every stone that is turned over turns out to hide new miracles, every path that one
takes leads to fascinating views. It sometimes seems as if String Theory solves
its own problems, and that we are just breathless spectators. Time after time
new contradictions appear to emerge, which however are always solved by a new
``String Miracle". It gives the theory an air of untouchability, and sometimes
makes you wonder how it could ever by falsified. 

These magical properties have a downside as well. They are fascinating
some theorists to such an extent that the ultimate goal is sometimes
forgotten. That goal now seems further away than in 1984. Despite great
promises, String Theory has not yet produced any measurable result. We
only see vague outlines of the Standard Model, and we are unable to work
out the details. Part of the reason is that we are still trying to
determine what String Theory really is.

Another reason is that we are faced with a huge number of possibilities.
Perhaps one would expect from a candidate ``Theory of Everything" that it is
unique, in other words, that there is only one possible outcome for the 
laws of physics. At present, however, the facts seem to point into a 
totally different direction.

When String Theory was reborn in 1984, there was a lot of talk about
``uniqueness", but that was clearly premature. In 1984 we did not know
one, but five String theories. Then things went out of control rapidly.
The five theories existed in a world with nine space and one time
dimensions, ten dimensions in total. Our space also has one time, but
only three space dimensions, height, width en depth. In the years after
1984 it became clear rapidly that it was much simpler to build
String Theories in four dimensions than in ten. Instead of five there
turned out to be billions of possibilities, actually infinitely many. 

I was involved in these developments in 1986, and in our paper we put
special emphasis on the large number of solutions. This was not regarded
as good news. Many years later I even met someone who claimed that our
work had convinced him to abandon String Theory and work on 
something else. Of course this was not our intention. I believe
I was one of the few people that regarded this large number of 
String Theories as a positive development, and I will explain in a 
moment why.

Gradually after 1986 some order was created in the chaos, and it started
becoming clear that all these distinct String Theories were really part
of a larger entity after all. They could all be related in one way or
another to one of the five String Theories in ten dimensions. To go from
ten to four dimensions, six of the ten dimensions are ``rolled up", so that
we can only observe four.

Usually we regard all these four-dimensional String Theories
as so-called ``ground states" of one of the ten-dimensional ones.
To understand the concept of a ground state one may think of a 
mountain range with many valleys. In this picture the mountain range
represents the theory, and the valleys the distinct ground states. Inhabitants
of each of these valleys only see a small part of the entire mountain
range. Nevertheless, everything is ultimately   connected.

This picture restores the lost uniqueness to some extent. There may be
many ground states, but there is just one theory. At least, that would be
the ideal situation. But up to now we still had five theories instead of one. 
That corresponds to five mountain ranges, each with its own valleys. Four
years ago this suddenly changed. The five mountain ranges all turned out to
be the same, but viewed from different angles. Just as the Mont Blanc looks
totally different from Italy or France, apparently totally different
String Theories turned out to be different ways of looking at the same
theory. Unfortunately we have not found an exact formulation of this
overarching theory. Many people are looking for that at the moment.

Everything seems to point in the direction that we are dealing with only
one theory, but that this theory has a huge number of ground states. Just
like each mountain valley may have its own laws and customs, every
ground state has its own system of laws of physics. Instead of electromagnetic,
strong and weak forces there will be other interactions, instead of quarks
and leptons other particles. If String Theory is correct, one of these
many possibilities is realized in our universe. One of the big challenges is
to demonstrate that our universe is indeed one of these possibilities.

We can only discuss other ground states purely theoretically. In contrast to
an inhabitant of a mountain valley, who in principle could go and have a look
at another valley, we would not even be able to exist in another ground state.
The quarks and leptons out of which we are composed do not even exist there.
Nevertheless it seems just a small step to assume that other ground states might be 
realized in another universe. Such a statement lies, however, beyond the
boundaries of physics. By definition, physics cannot make statements about
things that cannot be observed. We can only speak in theoretical terms about
other possible universes. They are solutions to the same equations satisfied
by our own universe. 

This line of thought fits in very well with a series of insights
that pointed out our modest place in the cosmos. Our planet is not
the center of the solar system, our sun is just one of many stars and
not even a very special one, and the same is true for our galaxy. It seems
natural to assume that also our universe, including the quarks, leptons
and interactions we observe is just one out of many possibilities.

This way of thinking has important consequences. If indeed our universe,
including its laws of physics and the entire Standard Model is just out one
of many possibilities, this implies that there are limits to what we
can compute. The properties of the quarks and leptons, their interactions and the
parameters of the Standard Model
(or at least part of them) were fixed at the birth of our universe, when a choice
 was made out of the many possibilities. We will never be able
to compute that choice, because it could just as well have been different.
 
I have the impression that many of my colleagues believe or hope that this
will ultimately not be the case. They hope to find a kind of mathematical
formula that has only one solution. That single solution should then correspond
to our four-dimensional world, including all quarks, leptons and the
four basic forces. Also the values of the nineteen (or more) parameters, such as
the masses of all particles, should then ultimately emerge as the outcome 
of a mathematical computation.

It could indeed be like that. At this moment it is not clear what 
will be left of this enormous number of ground states once we understand
String Theory properly. At first sight it may seem attractive that only
one should survive, but if you think about it for a moment it becomes
clear that this would  really be an undesirable end to the story.

I will try to explain this, but first I will make an instructive
historical comparison. Kepler, a great scientist whom we all know
because of his theory of planetary orbits, believed that the
distances of the planets to the sun were fundamental parameters, that
should be computable. He invented an ingenious theory with embedded
Platonic solids. With our current knowledge Kepler's attempt seems
absurd. Kepler did not know about the existence of the planets Uranus,
Neptune and Pluto, and this fact alone already destroys his theory. If he
had known about the existence of other solar systems he would not even have
started. Nevertheless one cannot blame Kepler for trying. One should not
assume too soon that something can never be computed. Once you have 
gathered enough insight it will become clear what can be computed and what cannot.
It seems very well possible to me that in some time we will arrive at the
insight that other worlds with other laws of physics are indeed possible.
Attempts to compute the parameters of the Standard Model will then
look equally naive as what Kepler had in mind.

Although the choice of distinct ground states is arbitrary from a 
mathematical point of view, it is not from a human point of view.
Our existence is tightly linked to the Standard Model and the
precise values of a number of parameters. A well-known example are
the masses of the two quarks out of which the proton and the neutron
are made. These quarks are called ``up" and ``down", and the latter is
heavier than the former. This has the pleasant consequence that the proton
is stable, whereas a free neutron decays in about fifteen minutes into
a proton and an electron. If we interchange the two quark masses it
is just the other way around. The proton would decay rapidly into a neutron and
a positron. The hydrogen atom would then be unstable, just as the water 
molecule. There is a long list of examples of this type. Some are more
impressive, but a bit harder to explain.

Our entire existence depends on a series of subtle processes that occurred 
during the evolution of the universe. These processes have finally led to
a planet where, for example, the crucial element Carbon occurs with
sufficient abundance. Various steps in this process depend critically
on the parameters of the Standard Model, such as the masses of particles and
the strength of interactions. It  seems often easy to demonstrate that even
small changes of certain parameters would obstruct the entire process.

From this point of view it would seem absurd that exactly those
parameter values would follow from a mathematical computation. We would be
left with a much bigger riddle than the one we are trying to solve. For this
reason I was very satisfied when it turned out that String Theory was
highly non-unique. If our planet were the only one in the Universe, it would
be a mystery why precisely that single planet would allow life. The fact
that there are billions of planets makes the mystery considerably less
severe. Analogously, the fact that many kinds of universes are possible makes
the existence of conditions for intelligent life in our universe considerably
less absurd than if there would be just one possibility.

This kind of reasoning is often referred to as the ``anthropic principle". In
reality this is a collective name for a variety of ideas, some of which I find
quite nonsensical. The anthropic principle states that the universe we 
observe is characterized by the fact that we exist as observers. It is
hard to disagree with this. One can, however, disagree with the consequences
that some people derive from it.

As far as I am concerned the anthropic principle only makes sense
if our universe it not the only possible one. Furthermore it only makes
sense within a completely consistent theory. Otherwise we do not know
which changes of the parameters of our universe make sense. For example, it
could well be that we are not allowed to change the up/down quark mass ratio at all.
Although I would like to have many options for the laws of nature of our
universe, I would still prefer the fundamental theory to be unique. String theory
seems to accommodate these two demands that at first sight seem contradictory:
according to our present insights there is just one theory, but many ground
states are possible. This is really the best outcome one could imagine.

The anthropic principle could be the explanation for certain coincidences
that are hard to understand from the point of view of the theory. It could
explain why up quarks are lighter than down quarks, but this mass difference
is not really that astonishing. It gets more interesting when we look at cases
where parameter ratios are very small or very large. Well-known examples
are the electron mass, which is improbably small, or the strength of the weak force.
The value of both parameters is essential for the evolution of our
universe and our existence, but this is not sufficient to conclude
that we understand the values of those parameters.

What really matters is the question if for other parameter values
some form of intelligent life is possible. If intelligent life would ultimately
develop in every imaginable universe, it would still be impossible to
understand why nature would go into extremes just to make our form of
intelligent life possible.

This leads to a formulation of the anthropic principle that might fall
within the boundaries of exact science. It requires first of all a theory
that we fully control, and of which we know all possible ground states, and
thereby all possible parameters and their allowed values. Furthermore we
need a definition of intelligent life that is not based on our
own quark-lepton world. For all the allowed ground states we should then
compute if intelligent life is possible. If this is only possible for
certain extreme parameter values, this would explain the values of those
parameters.

Unfortunately given our current knowledge
this programme is completely impossible to execute. It is already
impossible to say if life in our own universe is possible under
drastically changed conditions. It may seem reasonable to assume that
the element Carbon is required, but on the other hand it is very hard
to rule out other possibilities. I consider the question whether intelligent life is
possible in other valleys of the string mountain range extremely
interesting, but many times more difficult.
 
Nevertheless, for me a worthy end of the story of High Energy Physics
would already be the simple observation that we live in one of the many
valleys of the string mountain range. This is a modest, but perhaps just
achievable goal. If that is true, and if String Theory meets our
expectations we have a complete theory for everything that happens in our
universe, a theory that predicts precisely what we may expect from
future experiments, and that does not leave room for surprises. It would be
even better if we could also understand why we live in precisely this 
universe, and not in another. Unfortunately at this moment even the first,
modest goal is not more than a beautiful dream.

The foregoing was a sketch of a possible end of the story. It is the end
that given the current state of affairs seems the most desirable to me, but
in the end only hard results matter. Nature will probably not care much
about my wishes. Despite the word ``end" in the title it was
not at all my intention to suggest that the end will be reached soon. 
On the contrary, it will take many decades of work to produce a complete map of String Theory.
I am looking forward to an exciting continuation of this adventure. 

\section{The original Dutch text}

De Hoge Energie fysica neemt een speciale plaats in binnen
de natuurkunde. Dat blijkt eigenlijk al uit de naam.
De meeste vakgebieden binnen de natuurkunde zijn genoemd naar
wat zij bestuderen. De vaste stof fysica bestudeert eigenschappen
van vaste stoffen, de atoom fysica bestudeert atomen en de
kernfysica atoomkernen. Maar een Hoge Energie Fysicus bestudeert
geen Hoge Energie. De naam van het vak is ontleend aan de
experimentele technieken die gebruikt worden. Het wezen van het vak
is het bestuderen van de materie op steeds kleinere afstanden. 
Hiertoe worden bundels deeltjes met zeer hoge energie op elkaar
geschoten. Hoe hoger de energie, hoe verder men in de materie doordringt. 
Ons vak begint eigenlijk bij de uitvinding van de microscoop, maar
pas in deze eeuw is het echt goed op gang gekomen. We
hebben geleerd dat
materie bestaat uit moleculen, die weer zijn opgebouwd uit atomen.
De atomen bestaan uit een kern met daarom heen een wolk electronen,
en de kern bestaat weer uit protonen en neutronen. Steeds wanneer
je nog dieper kijkt lijken zich nieuwe verschijnselen voor te doen.
Wellicht is dat ook de reden dat het vak niet genoemd is naar het
onderwerp van studie; dat verandert immers steeds. Men spreekt 
weliswaar ook
wel van ``Elementaire deeltjes fysica", maar toen die naam voor
het eerst gebruikt werd dacht men nog dat protonen en neutronen
elementaire deeltjes waren. Intussen weten we wel beter. Het proton 
en het neutron zijn niet elementair, maar bestaan zelf weer uit andere
deeltjes, de quarks. Wanneer zou blijken dat quarks zelf ook weer
uit andere deeltjes zouden bestaan zou onze aandacht vanzelf verschuiven
naar nieuwe, nog diepere structuren. Voorlopig hebben we daar overigens geen
aanwijzigingen voor. 

Je zou dus kunnen zeggen dat
het wezenlijke verschil tussen de Hoge Energie Fysica en alle andere
takken van de natuurkunde is dat de eerste het zoekt in de diepte, terwijl
alle andere disciplines meer in de breedte werken. 
In die definitie van de Hoge Energie Fysica zit eigenlijk ook al
het einde van het vak opgesloten. Het lijkt
onvermijdelijk dat we eens tegen een grens zullen aanlopen die het
ons onmogelijk maakt om nog dieper in de materie door te dringen. 
Dit zou betekenen dat de natuur zich op het allerdiepste
nivo voor altijd aan onze waarneming onttrekt. 
Als wat we niet kunnen
zien net zo gecompliceerd is als wat we al gezien hebben, is er
geen enkele kans dat we het antwoord zouden kunnen raden. Dit lijkt
mij het meest waarschijnlijke einde van dit vak. 

Wanneer dat einde bereikt zal worden is minder duidelijk. 
De laatste 50 jaar zijn de technologische
grenzen steeds verder opgeschoven, en zijn er steeds indrukwekkender
machines gebouwd. Een van die successen is bijvoorbeeld LEP in Gen\`eve,
een ringvormige deeltjesversneller met een omtrek van 27 kilometer,
waarin electron en positron bundels met grote energie en verbijsterende 
precisie op elkaar geschoten worden. Steeds wanneer een dergelijke
machine opgeleverd wordt ligt de volgende alweer op de tekentafel. 
Technologisch gezien lijkt het einde misschien in zicht, maar
eigenlijk is dat al minstens twintig jaar het geval, en er zijn nu
nog steeds nieuwe versnellers gepland of in aanbouw. 

Een onvoorzienbare
technologische revolutie zou het einde van deze weg 
fors kunnen uitstellen, maar
uiteindelijk zullen we tegen fundamentele of practische grenzen
aanlopen. 
Dat wordt pijnlijk duidelijk wanneer we kijken naar de
kleinste afmetingen die volgens theoretici interessant zouden kunnen
zijn. In 
theorie\"en van Quantum Gravitatie spelen zich interessante verschijnselen
af op afstanden ter grootte van de Planck lengte, dat is ongeveer
$10^{-33}$ cm. Dit is zo vreselijk klein dat het waarnemen 
van dit soort structuren 
met onze krachtigste deeltjes-versnellers net zo iets is als
proberen protonen op de maan te zien met behulp van een telescoop.
Met een versneller als LEP kunnen we tot ongeveer een-honderdste van de
grootte van een proton doordringen. Een machine
die structuren ter grootte van de Planck lengte kan zien moet deeltjes
versnellen tot een gigantische Energie, de Planck Energie. 
Wanneer
we met de huidige technologie een dergelijke versneller zouden willen 
construeren,
dan blijkt dat die machine minstens 
de grootte van het gehele melkwegstelsel
zou moeten hebben. Het is een interessante vraag of dit zelfs
puur theoretisch mogelijk is, practisch mogelijk
is het in ieder geval niet. 

Hoewel de meerderheid van de experimenten met deeltjes-versnellers
gedaan wordt zijn er andere, wat minder directe methodes om iets 
over de diepere structuren van de natuur te weten te komen.
Maar welke methode ook gebruikt wordt, er zullen altijd   
verschijnselen denkbaar zijn die zich aan onze waarneming
onttrekken. We kunnen ons deeltjes voorstellen die dermate
weinig interactie met ons hebben, of gebeurtenissen die zo
onwaarschijnlijk zijn, dat we ze nooit zullen zien. 
Het is moeilijk in te zien waarom de natuur ons altijd
dermate vriendelijk gezind zou zijn dat we al haar geheimen 
uiteindelijk kunnen waarnemen.  

Dit vind ik een onplezierige gedachte. Ik zie de ontwikkeling van 
de fundamentele natuurkunde in deze eeuw als een spannend boek, dat je
moeilijk neer kunt leggen zonder het laatste hoofdstuk gelezen te hebben.
Het is een prachtig verhaal dat een waardig slot verdient. Het zou
een enorme teleurstelling zijn als het laatste hoofdstuk voor ons
onleesbaar zou blijken.

Wat is eigenlijk ons doel? Het echte antwoord op die vraag wordt
door de meeste van mijn collega's hoogstens met grote aarzeling
geformuleerd, omdat het enerzijds vrij arrogant en anderzijds
hopeloos naief klinkt. Je zou het kunnen omschrijven als 
de ``Theorie van Alles". Deze term werd enige jaren geleden voor
het eerst gebruikt in een vlaag van euforie over een nieuwe
doorbraak, en werd over het algemeen met hoongelach ontvangen.
Hoezo, Theorie van Alles?
Vertelt die theorie ons iets nieuws over supergeleiding?
Kun je er de genetische code van de
mens uit afleiden? Kun je er dan misschien het weer mee voorspellen?
De lijst van vragen waarmee mensen worstelen is eindeloos.  
Een theorie die al die 
vragen kan beantwoorden is ondenkbaar. 

Als we toch een bereikbaar en aanvaarbaar einddoel willen formuleren
zullen we de vragen waarop we een antwoord mogen
verwachten
moeten beperken. Dit houdt onder meer in dat de vragen niet te
ingewikkeld mogen zijn. Het DNA molecuul bijvoorbeeld
is dermate ingewikkeld,
dat het nooit direct uit een fundamentele theorie zal worden 
afgeleid. Niettemin weten we al decennia wat die fundamentele
theorie is: het is de Quantum Electrodynamica. Wanneer we voldoende
eenvoudige vragen stellen geeft deze theorie uitermate precieze
antwoorden. In de hoge energie fysica worden de meeste vragen 
uiteindelijk gereduceerd tot simple vragen zoals: wat gebeurt er precies
als twee deeltjes met een bepaalde energie op elkaar botsen. Wanneer
we ons ervan overtuigd hebben dat we alle vragen van dat type kunnen
beantwoorden is wat mij betreft 
aan een van de eisen voor een Theorie van Alles voldaan.
  
Eigenlijk hebben we al bijna een dergelijke theorie tot onze
beschikking. Deze wordt aangeduid met de weinig indrukwekkende 
naam ``Standaard Model". Deze theorie beschrijft drie van de
vier fundamentele krachten die we kennen. Op de eerste plaats is
dat de al genoemde Quantum Electrodynamica. Deze interactie
manifesteert zich op vele manieren in de wereld om ons heen:
licht, electriciteit, magnetisme, chemie, en de meeste eigenschappen van
materialen vallen allemaal onder deze theorie. Dan is er
de Quantum Chromodynamica, die er onder andere voor zorgt dat de
protonen en neutronen in kernen bij elkaar gehouden worden. Tenslotte
is er de zwakke kracht, die zich 
bijvoorbeeld manifesteert in radioactiviteit. Voor al deze 
verschijnselen hebben we een uitermate succesvolle fundamentele
theorie, die al vijfentwintig jaar lang een indrukwekkende reeks
experimentele verificaties met verve heeft doorstaan. Wanneer
men in de toekomst terugkijkt op deze eeuw zal het Standaard Model
ongetwijfeld genoemd worden als een van de grote doorbraken in
de fysica, samen met de quantum mechanica en de relativiteits theorie.

Toch zal niemand geneigd zijn de 
weinig pretentieuze naam ``Standaard Model" te
vervangen door ``Theorie van Alles". Daar zijn een aantal redenen voor.
Op de eerste plaats ontbr\'e\'ekt een van de vier bekende
interacties, namelijk de zwaartekracht. Dat lijkt 
op het eerste gezicht geen al te groot
probleem. Ook voor de zwaartekracht hebben we een uitstekende
theorie, Einstein's algemene relativiteits theorie, een
van de parels van de natuurkunde van deze eeuw. 
Net als
het Standaard Model geeft die theorie uiterst nauwkeurige
voorspellingen die in overeenstemming zijn met een groot aantal
experimenten. Helaas leeft deze theorie echter op gespannen voet
met de quantum mechanica. Concreet betekent dit dat we geen 
voorspellingen kunnen doen wanneer
deeltjes met Planck energie op elkaar botsen, of voor extreme
verschijnselen in de buurt van zwarte gaten.
Het is uitermate onwaarschijnlijk dat we dit soort processen ooit
daadwerkelijk kunnen waarnemen, maar van een Theorie van Alles
verwachten we dat deze ook in dergelijke extreme omstandigheden
een antwoord geeft. 

Het Standaard Model heeft daarentegen geen problemen met
de quantum mechanica. Het is geformuleerd
in termen van Quantum Velden Theorie, en rust stevig op zijn
twee pijlers, de quantum mechanica en de relativiteits theorie.
Toch zijn we er allerminst
gerust op dat het Standaard Model geheel in orde is. 
Het Standaard Model bevat 17 deeltjes, waarvan we er tot nu toe
16 gevonden hebben. Er ontbreekt er dus nog een, het
zogenaamde Higgs deeltje. Dit deeltje verschilt nogal van 
de andere 16, en het wordt algemeen gezien als de zwakste
plek in het Standaard Model. 
Het bestaan van dit deeltje is al vijfentwintig jaar geleden
gepostuleerd, omdat het essentieel is voor de
consistentie van het Standaard Model. Maar anderzijds werd al
snel duidelijk dat het een nieuw probleem introduceert. 
Wat dat probleem is zal ik hier niet in detail uitleggen. Het komt
er op neer dat de zwakke kracht, ondanks zijn naam, eigenlijk vele
malen sterker is dan je redelijkerwijs zou mogen verwachten.
 
Er is nog meer wat ons niet bevalt aan het Standaard Model. 
Zo heeft het bijvoorbeeld wat veel parameters, namelijk minstens
19 en waarschijnlijk meer. Een parameter is een getal waarvan de
waarde niet door de theorie wordt vastgelegd. 
Zo vertelt het Standaard Model ons bijvoorbeeld niet hoeveel
de quarks wegen. De massa's van de zes quarks kunnen alleen
bepaald worden door ze te meten. Wanneer alle parameters gemeten
zijn ligt de theorie vast en kan elke andere grootheid berekend 
worden. Hoewel 19 parameters niet echt veel is hadden we er misschien
liever wat minder, bijvoorbeeld geen enkele.  

Wat ons echter nog meer zorgen baart is het feit dat sommige van 
die parameters nogal speciale waarden hebben. Zo is bijvoorbeeld 
de verhouding van de electron massa met die van de zwaarste quark,
de top quark, heel erg klein, minder dan een honderdduizendste.
Aangezien het Standaard Model niets zegt
over de massa's, had er elk getal uit kunnen komen. 
Als je dan constateert
dat in de natuur zo'n getal heel erg klein is krijg je het gevoel
dat je dit zou moeten kunnen begrijpen. 
Het eerder genoemde probleem met het Higgs boson is eigenlijk van
dezelfde aard, maar nog wat erger.
Het is net als wanneer je je in een
zaal met honderd mensen bevindt, en je ontdekt dat ze allemaal op
dezelfde dag jarig zijn. De kans dat dit toeval is, is dermate klein, dat
je meteen aanneemt dat er een verklaring
voor moet zijn. In de natuurkunde van ons heelal 
zien we verscheidene van dergelijke onwaarschijnlijke toevalligheden. 

Iets anders wat velen niet bevalt is dat het Standaard Model bij lange na
niet de enige mogelijkheid is. Toen Einstein de 
Algemene Relativiteits theorie opschreef had hij eigenlijk niet veel 
keus. Maar de bedenkers van het Standaard Model hadden talloze
variaties voorhanden. Dertig jaar intensief experimenteel werk hebben
vrijwel al deze variaties naar de prullenmand verwezen, maar eigenlijk
zouden velen dat liever op puur theoretische gronden willen kunnen doen.
Men zou dus liever een theorie hebben die uniek is.    

Niettemin is dit een soort vragen waarvan het wat minder zeker
lijkt dat ze beantwoord moeten kunnen worden.
Eigenlijk hebben we in het Standaard Model alleen nog
vragen over van dit type. Het zijn allemaal vragen die met ``waarom"
beginnen: waarom juist dit Standaard Model,  
waarom hebben de parameters de waarden
die we waarnemen, en waarom zijn sommige waarden zo klein.   
Het is onvermijdelijk dat er in een theorie vragen van dit type
overblijven. Het feit dat dergelijke vragen niet beantwoord worden
is dus geen reden om aan een theorie het predicaat ``Theorie van Alles"
te onthouden.  Dat zouden we wel moeten doen als het inconsistent was
of in strijd met experimenten. 

Wat ik met het voorgaande heb willen zeggen is dat het begrip
``Theorie van Alles" niet noodzakelijk een luchtkasteel is, mits
men er niet de verkeerde verwachtingen van heeft. Het Standaard Model is
niet alleen een uitstekend model voor drie van de vier natuurkrachten
en alle materie, maar ook
voor het concept ``Theorie van Alles". Op een paar punten na is
het eigenlijk precies waar we naar op zoek zijn.    

Het Standaard Model heeft een ander probleem dat zelden
als zodaning genoemd wordt, maar voor mij wel essentieel is. 
Het is weliswaar volledig
in overeenstemming met alles wat we kunnen waarnemen, maar het heeft
niets te zeggen over wat we niet kunnen zien. 
Wanneer nieuwe experimenten nieuwe deeltjes of nieuwe
interacties ontdekken zullen we het Standaard Model moeten uitbreiden.
De Quantum Velden Theorie laat tal van uitbreidingen toe. We kunnen
wel postuleren dat de natuur beschreven wordt door het Standaard Model
zonder enige toevoeging, maar niets garandeert dat. We zijn dus altijd
afhankelijk van verdere experimenten. 
Dit lijkt zo vanzelfsprekend dat eigenlijk niemand het ooit als
probleem noemt. 

Hoe staat het nu met de kansen om een aanvaardbare theorie ook 
werkelijk te vinden? Ik heb al gezegd dat ik daar niet optimistisch
over ben. Het Standaard Model is niet zo vreselijk ingewikkeld. Als je
de juiste taal spreekt kun je het in twee of drie regels opschrijven.
Niettemin zijn er vele experimenten nodig geweest om zover te komen.
Experiment en Theorie gaan nu eenmaal hand in hand. 
Zonder experimentele hulp maken we weinig kans.

Toch is er misschien nog een kleine kans.
Er is op dit moment binnen de theorie een ontwikkeling
gaande die ons enige hoop geeft dat we onze taak misschien
toch tot een goed einde kunnen brengen. 
Ik doel hier op een iets wat we gewoonlijk ``String Theorie" noemen.
In het Nederlands wordt dat gewoonlijk met ``Snaar Theorie" vertaald, maar
ik vind dat niet prettig klinken. De theorie ontleent zijn naam
uit het feit dat alle deeltjes gerealizeerd zijn als trillingen
van een soort fundamentele 
snaar. De grondtonen corresponderen met de deeltjes
van het Standaard Model, plus misschien een aantal nieuwe deeltjes.
 De boventonen
corresponderen met een oneindige reeks deeltjes die we nooit zullen
waarnemen, tenzij we een versneller met Planck energie zouden kunnen bouwen. 
Mijn voorkeur voor het woord ``String" in plaats van ``Snaar" heeft
vermoedelijk te maken met de ruimere betekenis van het Engelse woord,
wat immers ook koord of touwtje kan betekenen. Die betekenis past 
beter bij het beeld wat ik gedachten heb dan een strak gespannen viool
snaar.

Overigens is intussen wel duidelijk geworden dat het woord ``String"
veel te beperkt is en het woord ``Theorie" een beetje voorbarig. 
We hebben ontdekt dat naast snaren ook membranen een rol spelen, en
we hebben ons ook gerealizeerd dat wat we tot nu toe ``String Theorie"
noemden slechts een klein hoekje is van een werelddeel wat 
nog grotendeels in kaart gebracht moet worden. Bij gebrek aan een
betere naam zal ik dat hele werelddeel voorlopig maar ``String Theorie"
noemen. 

String Theorie werd dertig jaar geleden min of meer bij toeval
ontdekt bij pogingen om de sterke kracht te begrijpen. Al snel
werd duidelijk dat deze theorie over magische eigenschappen beschikte,
maar voor de beschrijving van de sterke kracht werd al spoedig
een beter alternatief gevonden, de Quantum Chromodynamica. Tien
jaar lang leidde de theorie, samen met een handjevol trouwe
volgelingen, een kwijnend bestaan, om in 1984 een triomphale
terugkeer te beleven. Inmiddels was duidelijk geworden dat de
theorie veelbelovende eigenschappen had voor de beschrijving
van een heel andere kracht, de zwaartekracht. Enige verlichte
geesten hadden zich dat overigens in 1975 al gerealizeerd, maar
zij vonden toen weinig gehoor. 

De revolutie van 1984 is vooral te verklaren uit het feit
dat String Theorie ons iets cadeau gaf waar eigenlijk niemand op
gerekend had. Naast een mogelijk consistentie theorie van quantum
gravitatie bleek String Theorie ook deeltjes en
krachten te bevatten die leken op het Standaard Model. Dit is een
algemeen kenmerk van goede theorie\"en: slechte theorie\"en komen
altijd wat te kort, maar goede theorie\"en geven je van alles
cadeau waar je redelijkerwijs niet op mocht rekenen.  

Als alles uitkomt zoals we hopen zou dit uiteindelijk
een theorie moeten opleveren die geheel consistent is en alle
bekende deeltjes en krachten bevat. Eerder heb ik al gezegd dat ik
eigenlijk ook zou willen dat een fundamentele theorie een volledige
beschrijving geeft van alles wat we (nog) niet kunnen waarnemen.
String Theorie heeft een 
eigenschap die dit lijkt te verwezenlijken. Ik doel hier op 
iets dat in speciale gevallen
met de technische term ``Modulaire Invariantie" wordt
aangeduid, en waarvan de precieze vorm in algemene zin nog niet
duidelijk is.  In ieder geval wijst alles er op
dat je aan een gegeven String Theorie
niet straffeloos deeltjes kunt toevoegen of deeltjes kunt weglaten.    
Als we dus eenmaal een String Theorie hebben gevonden die exact
het Standaard Model bevat, kun je daar
niets meer aan toevoegen. Aan een String Theorie kan niet gesleuteld worden. 
Het is best mogelijk dat de theorie nieuwe deeltjes voorspelt, maar als
een experiment een niet verwacht
deeltje vindt, dan stort daarmee het gehele bouwwerk in. String Theorie
kan dus eigenlijk 
alleen maar een Theorie van Alles opleveren, of we die term
nu prettig vinden of niet: Het is Alles of Niets.

Het woord ``Niets" is een beetje overdreven. Op zijn minst hebben
we een adembenemend stukje wiskunde in handen.
String Theorie heeft een grote
aantrekkingskracht op theoretici doordat het over
wonderbaarlijke eigenschappen beschikt die aan buitenstaanders
wat moeilijk zijn uit te leggen. Het is een reusachtig landschap 
waar theoretici zich doorheen bewegen met een soort Alice-in-Wonderland
gevoel. Elke steen die omgekeerd wordt blijkt nieuwe wonderen te
verbergen, elk pad dat ingeslagen wordt leidt tot fascinerende
vergezichten. Het lijkt soms of String Theorie zijn eigen problemen
oplost, waarbij wij slechts ademloze toeschouwers zijn. Keer op keer
lijken zich interne tegenstellingen te openbaren, die echter steevast
door een nieuw ``String Wonder" opgelost worden. Het geeft de theorie
iets onaantastbaars, en je vraagt je soms af hoe String Theorie
nog gefalsificeerd zou kunnen worden. 

Die magische eigenschappen hebben ook een schaduwzijde. Ze fascineren
sommige theoretici zodanig dat het uiteindelijke doel soms uit het oog
wordt verloren. Dat doel lijkt nu verder weg dan in
1984. 
Ondanks de mooie beloftes heeft 
String Theorie nog geen enkel meetbaar resultaat opgeleverd.
Van het Standaard Model zien we hoogstens een vage schim, de details
kunnen we nog niet uitwerken.  
Dat komt voor een deel omdat we nog steeds bezig zijn te ontdekken
wat String Theorie eigenlijk is. 

Een andere reden is dat we met vreselijk veel mogelijkheden te maken
hebben.
Men zou misschien van een kandidaat 
Theorie van Alles verwachten dat deze eenduidig is, met andere
woorden, dat er maar \'e\'en mogelijke uitkomst is voor de wetten van de
natuur. Voorlopig lijken
de feiten echter een geheel andere richting uit te wijzen.

Toen String Theorie in 1984 herboren werd, werd veel over ``uniciteit"
gesproken, maar dat was duidelijk voorbarig. Ik 1984 kenden we namelijk
niet \'e\'en, maar vijf String Theorie\"en. Daarna liep het snel uit de hand.
Die vijf String Theorie\"en bestonden namelijk in een wereld met negen
ruimte en \'e\'en tijd dimensie, dus tien dimensies in totaal. Onze eigen 
ruimte heeft ook \'e\'en tijd, maar slechts drie ruimte dimensies,
corresponderend met hoogte, breedte en diepte. In de jaren na 1984 
werd snel duidelijk dat het in vier dimensies veel eenvoudiger
was om String Theorie\"en te construeren dan in tien. In plaats van vijf
bleken er miljarden mogelijkheden te zijn, of eigenlijk oneindig veel.

Ik was in 1986 zelf bij die ontwikkelingen betrokken, en in ons artikel
werd dat grote aantal oplossingen met enige nadruk genoemd. Dat werd
niet als goed nieuws beschouwd. Vele jaren later heb ik zelfs iemand
ontmoet die beweerde dat ons werk hem ervan had overtuigd String Theorie
te verlaten en iets anders te gaan doen. Dat was uiteraard niet onze 
bedoeling. Zelf was ik, denk ik, een van de weinigen die dit grote
aantal String Theorie\"en w\'el als een positieve ontwikkeling beschouwde,
en ik zal later uitleggen waarom.    

Geleidelijk kwam er na 1986 wat orde in de chaos, en het begon
duidelijk te worden dat al die verschillende String Theorie\"en eigenlijk
toch deel uitmaakten van een groter geheel. 
Ze konden allemaal
op een of andere manier 
gerelateerd worden met een van de vijf String Theorie\"en in tien
dimensies.
Om van tien naar vier dimensies te komen worden 
dan zes van de tien dimensies als het ware
opgerold, zodat wij er slechts vier kunnen waarnemen. 

Gewoonlijk beschouwen we al die vier-dimensionale String Theorie\"en
als zogenaamde ``grondtoestanden" van een van de vijf tien-dimensionale. 
Om het begrip grondtoestand te begrijpen kan men denken aan een gebergte
met vele dalen. In dit beeld correspondeert het gebergte
met de theorie, en de verschillende dalen met de verschillende
grondtoestanden. De bewoners van ieder van die dalen
zien slechts een klein deel van het gehele gebergte. Niettemin is
uiteindelijk alles met elkaar verbonden.

Dit beeld herstelt de verloren eenduidigheid weer enigszins. Weliswaar
zijn er vele grondtoestanden, maar er is slechts \'e\'en Theorie.  
Althans, dat zou het ideaal zijn. We hadden echter nog steeds vijf
theorie\"en in plaats van \'e\'en. Dat correspondeert dus met vijf
gebergtes, ieder met zijn eigen dalen.
Vier jaar geleden kwam hier plotseling verandering in.
De vijf gebergtes bleken in werkelijk allemaal hetzelfde, maar vanuit
verschillende hoek bekeken. Net zoals de Mont Blanc er vanuit 
Frankrijk gezien geheel anders uit ziet dan vanuit Itali\"e, zo bleken
ook op het oog totaal
verschillende String Theorie\"en eigenlijk verschillende
benaderingen van dezelfde theorie te zijn.  
Helaas hebben we nog geen
exacte formulering van die overkoepelende theorie. 
Daar wordt op dit moment druk naar gezocht.

Alles lijkt er op dit moment op te wijzen dat we weliswaar te maken
hebben met \'e\'en theorie, maar die theorie heeft een enorm aantal
grondtoestanden.
Net zoals elk bergdal vaak zijn
eigen wetten en gebruiken kent, zo behoort bij elke grondtoestand
een ander stelsel natuurwetten.
 In plaats van electromagnetische, sterke en zwakke
krachten zullen er andere interacties zijn, in plaats van quarks
en leptonen andere deeltjes. 
Als String Theory correct is, is in ons heelal   
\'e\'en van die vele mogelijkheden gerealizeerd. Een
van de grote uitdagingen is aan te tonen dat ons heelal inderdaad
tot de mogelijke oplossingen behoort.

Over andere grondtoestanden kunnen we slechts puur theoretisch discussieren.
In tegenstelling tot een bewoner van een bergdal, die in principe
een ander dal zou kunnen gaan bekijken, kunnen wij in een andere
grondtoestand niet eens bestaan. De quarks en electronen waar wij uit
zijn opgebouwd bestaan daar immers niet eens. Toch lijkt het slechts
een kleine stap om nu aan te nemen dat andere grondtoestanden wellicht
in een ander heelal gerealizeerd zouden kunnen zijn. Een dergelijke
bewering ligt echter buiten de grenzen van de natuurkunde. Per definitie
kan de natuurkunde geen uitspraken doen over zaken die niet
waargenomen kunnen worden. We kunnen over andere mogelijke ``heelallen"
slechts theoretisch spreken. Het zijn oplossingen van 
dezelfde vergelijkingen waar ook ons heelal aan voldoet. 

Deze gedachtengang past goed in een reeks van inzichten die
ons op onze bescheiden plaats in de cosmos hebben gewezen. 
Onze planeet bleek niet het centrum van het zonnestelsel, onze
zon is slechts een van de vele sterren en niet eens een heel bijzondere,
en hetzelfde geld voor ons gehele melkwegstelsel. Het lijkt vrij
natuurlijk om aan te nemen dat ook ons heelal, inclusief de quarks,
leptonen en interacties die we waarnemen slechts een van vele mogelijke
is.     

Er zijn belangrijke consequenties verbonden aan deze 
manier van denken. Als het inderdaad zo is dat ons heelal, inclusief
zijn natuurwetten en het gehele Standaard Model slechts een van vele
mogelijkheden is, dan impliceert dit dat er grenzen zijn aan wat we
kunnen uitrekenen. De eigenschappen van de quarks en leptonen, hun
interacties, en de parameters van het Standaard Model (of althans
een deel daarvan) zijn bij het ontstaan van het heelal vastgelegd,
waarbij een keuze gemaakt is uit de vele mogelijkheden. Die keuze
zullen we nooit kunnen uitrekenen,
want die had immers ook anders kunnen zijn. 

Ik heb de indruk dat vele van mijn collega's geloven of hopen dat dit
uiteindelijk niet het geval  
zal zijn. Men hoopt
een soort wiskundige formule te vinden die maar \'e\'en uitkomst heeft.
Die ene uitkomst zou dan moeten corresponderen met onze 
vier-dimensionale wereld, 
inclusief alle quarks, leptonen en de vier basiskrachten. Ook de 
waarden van de negentien (of meer)
parameters, zoals de massa's 
van alle deeltjes zouden
uiteindelijk als oplossing
uit een wiskundige berekening moeten komen rollen.

Het zou inderdaad zo kunnen zijn.
Op dit moment is niet duidelijk wat er van die enorme 
aantallen
grondtoestanden overblijft wanneer we String Theory echt goed
begrijpen. 
Het lijkt op het
eerste gezicht misschien aantrekkelijk
dat er maar \'e\'en overblijft, maar als je er even over nadenkt 
wordt duidelijk dat dit
eigenlijk een ongewenst einde van het verhaal zou zijn. 

Ik zal
proberen dat duidelijk te maken, maar eerst wil ik
een leerzame historische vergelijking maken. 
Kepler, een groot geleerde die we allemaal kennen van zijn
theorie van de planetenbanen, dacht dat de 
afstanden van de planeten tot de zon fundamentele
parameters waren, die uitgerekend moesten kunnen worden.
Hij bedacht er een ingenieuze theorie voor met in elkaar
ingebedde regelmatige veelvlakken. 
Met onze huidige kennis van zaken lijkt Kepler's poging absurd.
Kepler wist niet van het bestaan van de planeten Uranus, Neptunus en Pluto,
en dit feit alleen al haalt zijn theorie onderuit.
Als hij geweten had van het bestaan van andere zonnestelsels
was hij er niet eens aan begonnen.  
Toch kun je Kepler niet verwijten dat hij het probeerde. Je moet niet te
snel aannemen dat iets nooit uitgerekend zal kunnen worden. Wanneer
je voldoende inzicht verwerft zal uiteindelijk vanzelf wel duidelijk
worden wat je wel en niet kunt uitrekenen.
Het lijkt me heel goed mogelijk dat we 
over enige tijd tot het inzicht komen dat er inderdaad 
ook andere werelden met andere natuurwetten mogelijk zijn. Pogingen om alle
parameters van het Standaard Model uit te rekenen zullen dan 
even naief overkomen als wat Kepler beoogde.  
 
Hoewel de keuze uit de verschillende grondtoestanden 
willekeurig is vanuit wiskundig oogpunt, is dat niet zo vanuit
menselijk oogpunt. Ons bestaan is zeer nauw verbonden met het
Standaard Model en de precieze waarden van een aantal parameters.
Een bekend voorbeeld zijn de massa's van de twee quarks waaruit
het proton en het neutron bestaan. 
Die quarks worden ``up" en ``down" genoemd, en het tweede is zwaarder
dan het eerste. 
Dit 
heeft het plezierige gevolg dat 
het proton
stabiel is, terwijl een vrij neutron in ongeveer 15 minuten uiteen valt in
een proton en een elektron. Wisselen we de twee quark massa's om
dan is het net andersom. Het proton zou dan
zeer snel vervallen in een neutron en een positron. Het waterstof
atoom zou niet stabiel zijn, evenmin als het water molecuul. Ons leven
zou volstrekt onmogelijk worden, we zouden niet eens ontstaan zijn.
Er bestaat een  lange lijst van voorbeelden van dit type. 
Een aantal
daarvan is nog indrukwekkender, maar wat lastiger uit te leggen. 

Ons gehele bestaan hangt af van een subtiele reeks processen die
zich gedurende de ontwikkeling van ons heelal hebben afgespeeld. 
Die processen hebben uiteindelijk een planeet opgeleverd waar
bijvoorbeeld het cruciale element koolstof in voldoende mate
aanwezig is. Verscheidene stappen in dit process hangen op
een kritieke manier af van parameters in het Standaard Model, 
zoals de massa's van deeltjes en de sterktes van interacties. Het lijkt
vaak eenvoudig aan te tonen dat zelfs een vrij kleine wijziging van 
bepaalde parameters het proces volledig lam gelegd zou hebben. 

Vanuit dit oogpunt lijkt het absurd dat precies die parameter
waarden uit een wiskundige berekening zouden volgen. We zouden dan met
een veel groter raadsel blijven zitten dan we proberen op te lossen.
Om deze reden was ik ook zeer tevreden toen bleek dat String Theorie in
hoge mate niet uniek was. Wanneer onze planeet de enige in het heelal
zou zijn, zou het een raadsel zijn waarom juist die ene planeet leefbaar
was. Het feit dat er miljarden planeten zijn maakt
het raadsel aanzienlijk minder ernstig. Op analoge wijze maakt het feit
dat er vele soorten heelal mogelijk zijn, het bestaan van
condities voor intelligent leven in \'ons heelal aanzienlijk minder
absurd dan wanneer er slechts \'e\'en mogelijkheid zou zijn. 

Dit soort redeneringen  
wordt vaak aangeduid met de term
``anthropisch principe".  Dit is eigenlijk
een verzamelnaam voor verscheidene  idee\"en, 
waarvan ik er een aantal
overigens vrij onzinnig vind. Het anthropisch
principe stelt dat het heelal dat wij 
waarnemen wordt gekenmerkt door het feit dat wij er zijn om het
waar te nemen. Hiermee kan 
eigenlijk niemand het oneens zijn. Wel met de consequenties
die sommigen hieraan verbinden.  

Het anthropisch principe heeft wat mij betreft alleen zin als
ons heelal niet het enig mogelijke is. Het heeft bovendien alleen
zin binnen een geheel consistente theorie. Anders weten we namelijk niet
welke veranderingen van de parameters van ons heelal zinvol zijn. Het zou
bijvoorbeeld best zo kunnen zijn dat we de up/down quark massaverhouding
helemaal niet kunnen veranderen. Hoewel ik graag vele mogelijkheden
voor de
natuurwetten van het heelal zou willen hebben, zou ik ook graag zien 
dat de fundamentele theorie uniek was. 
String Theorie lijkt die twee op eerste gezicht 
tegenstrijdige eisen in zich te verenigen: volgens onze huidige
inzichten is er slechts \'e\'en theorie, maar zijn er vele grondtoestanden
mogelijk. Dit is eigenlijk de best denkbare uitkomst. 

Het anthropisch principe zou de verklaring kunnen 
geven voor bepaalde toevalligheden die vanuit het oogpunt
van de theorie moeilijk te begrijpen zijn. 
Het zou kunnen verklaren waarom up quarks lichter zijn dan 
down quarks, maar echt verbazingwekkend is dit massaverschil
eigenlijk niet. Het wordt interessanter wanneer we kijken naar
gevallen waar parameter verhoudingen zeer klein of zeer groot zijn.
Bekende voorbeelden zijn de electron massa, die onwaarschijnlijk
klein is, of de sterkte van de zwakke kracht.
De waarde van beide parameters is essentieel
voor de ontwikkeling van ons heelal en ons bestaan, maar 
dit dit is niet voldoende om te concluderen dat we de waarden 
van die parameters begrijpen.

Waar het werkelijk om gaat is de vraag of er voor 
andere parameter waarden enige vorm van intelligent leven mogelijk is.
Als intelligent leven zich in elk denkbaar heelal uiteindelijk
zou ontwikkelen is het nog altijd niet te begrijpen waarom de natuur zich
in extremen zou begeven om juiste onze vorm van intelligent leven
mogelijk te maken. 

Dit leidt tot een
formulering van het anthropisch principe die wellicht binnen het
kader van de exacte wetenschap valt. Het vereist op de eerste plaats
een theorie die we geheel onder controle hebben en waarvan we
in het bijzonder alle mogelijke grondtoestanden kennen, en daarmee alle mogelijke parameters en hun toegestane 
waarden. Verder hebben we een definitie van intelligent leven nodig
die niet op onze eigen quark-lepton wereld gebaseerd is. Voor elk van 
de toegestane grondtoestanden zouden we dan moeten uitrekenen of
intelligent leven mogelijk is. Als dit alleen mogelijk is voor
bepaalde extreme parameter waarden, zijn die waarden daarmee 
verklaard.  

Helaas is dit programma met onze huidige kennis van zaken
volstrekt onuitvoerbaar. Het is al vrijwel onmogelijk om
te zeggen of leven in ons eigen heelal mogelijk is onder
drastisch gewijzigde condities. Het lijkt bijvoorbeeld redelijk
om aan te nemen dat hiervoor het element koolstof vereist is, maar 
het is anderszijds bijzonder moeilijk om andere mogelijkheden
uit te sluiten. Of er in een van de andere dalen van het 
String gebergte intelligent leven kan bestaan is een vraag die ik
bijzonder interessant, maar vele malen moeilijker vind. 

Hoe het ook zij, een waardig slot van het verhaal van de
Hoge Energie fysica zou voor mij al de eenvoudige constatering
zijn dat wij in een van die vele dalen van dat String gebergte
leven. Dit is een bescheiden, maar heel misschien haalbaar doel.
Als dat zo is, en als String Theory aan onze
verwachtingen voldoet, hebben we een complete theorie voor alles
wat er in ons heelal gebeurt, een theorie die precies voorspelt wat we in
eventuele toekomstige experimenten nog mogen verwachten, en die
geen ruimte laat voor verrassingen.  
Het zou helemaal
mooi zijn als we ook nog kunnen begrijpen waarom wij
juist in dit heelal leven 
en niet in een ander. Helaas is op dit moment zelfs het eerste,
bescheiden doel niet meer dan een mooie droom.  

Het voorgaande was een schets van een 
mogelijke afloop van het verhaal.
Het is de afloop die
mij gezien de huidige
stand van zaken
het meest wenselijk lijkt, maar uiteindelijk tellen
slechts de harde resultaten.
De natuur zal zich van mijn wensen vermoedelijk weinig aantrekken. 
Ondanks het woord ``slot" in de titel
was het bepaald niet mijn bedoeling om te suggeren dat het einde
al spoedig bereikt zal worden. Integendeel, het zal nog vele decennia
werk vergen om String Theorie volledig in kaart
te brengen. Ik verheug me op een boeiende voortzetting van dit avontuur.
\newpage

\section{Added Notes}

After posting the first version of this paper I received several reactions
pointing out related remarks and ideas. I decided not to modify the original 
text (apart from correcting misprints), but to  add this section.

Paul Townsend sent me a response to the footnote on page 2
and suggested the first paragraph of \cite{Gibbons:1993sv} as an example, which indeed
it is. Lee Smolin drew my
attention to  his
ideas on ``cosmological natural selection" \cite{Smolin:1994vb}. I knew this
work, but not that his ideas originated from worries about the string vacuum
 explosion, as explained in chapter 5 of his book \cite{Smolin:1997rs}. This chapter 
 certainly needs to be mentioned in this context, although Smolin's point of
 view is rather different from mine. He also pointed out Strominger's 1986 paper
 \cite{Strominger:1986uh}, presumably  one of the very first to raise
 the possibility of a highly non-unique outcome.  Other papers from 1986 that made this point
 were the  ones on the fermionic constructions,  
 \cite{Kawai:1986va} and  \cite{Antoniadis:1986rn} and 
 the bosonic construction \cite{Lerche}, already mentioned in the original text. It is
 rather interesting to see how the various authors commented on this issue. 
 
I wanted to avoid discussing the vast number of anthropic ideas and their history,
and consequently did not include any references to papers in this area,
with the exception of what I  called ``Weinberg's anthropic argument" for 
the cosmological constant. However, as John Barrow correctly pointed out,
an earlier (and different) anthropic argument for a small cosmological constant
appeared in the book by Barrow and Tipler \cite{BT}, which contains an 
extensive discussion of anthropic ideas, and many references. Now that I
have entered into the tricky business of citing anthropic papers, 
I should definitely mention the ideas of Andrei Linde on the anthropic
principle in inflationary cosmology, going back to 1983 (see \cite{Linde:2002gj},
and references therein).  I learned about his work
a few years later, when I started discussing my point of view with some colleagues
at CERN\rlap.\footnote{The way I learned about his work was that someone at CERN told me
``what you are saying sounds like this disgusting anthropic principle that Linde is
writing about". I have never understood the reason for the disgust.} However, I do not recall
any specific paper that influenced my thoughts on the anthropic principle. I am certain
that I knew about the up/down quark mass interchange argument, and that I had seen
the term ``anthropic principle" before 1986, but  I do not even remember if I had any opinion on it. 
It was really the apparent non-uniqueness of
the string vacuum that forced me in this direction.

\vskip 1.5truecm

\bigskip

{\bf Acknowledgements:}
I would like to thank Beatriz Gato-Rivera for carefully reading the English part of the text, and
for comments and discussions.
Furthermore I would like to thank all people who reacted for their appreciative
comments, and especially L. Susskind for his enthusiastic and generous message.

\break

\end{document}